\documentclass[lettered]{aastex}

\slugcomment{Based on observations with the Las Campanas Observatory Magellan Telescopes.}

\shorttitle{Discovery of a jet from Sanduleak's star in the LMC}
\shortauthors{Angeloni et al.}

\begin{document}

\title{Discovery of a giant, highly-collimated jet \\from Sanduleak's star in the Large Magellanic Cloud}

\author{R. Angeloni}
\affil{Departamento de Astronom\'ia y Astrof\'isica, Pontificia Universidad Cat\'olica de Chile, Vicu\~{n}a Mackenna 4860, 7820436 Macul, Santiago, Chile}
\email{rangeloni@astro.puc.cl}

\author{F. Di Mille}
\affil{Australian Astronomical Observatory - Carnegie Observatories\\ Colina El Pino, Casilla 601, La Serena, Chile}
\email{fdimille@lco.cl}

\author{J. Bland-Hawthorn}
\affil{Sydney Institute for Astronomy, School of Physics, University of Sydney\\ NSW 2006 Sydney, Australia}
\email{jbh@physics.usyd.edu.au}

\and

\author{D. J. Osip}
\affil{Las Campanas Observatory, Carnegie Observatories\\ Colina El Pino, Casilla 601, La Serena, Chile}
\email{dosip@lco.cl}

\begin{abstract}
Highly-collimated gas ejections are among the most dramatic structures in the Universe, observed to emerge from very different astrophysical systems - from active galactic nuclei down to young brown dwarf stars. Even with the huge span in spatial scales, there is convincing evidence that the physics at the origin of the phenomenon, namely the acceleration and collimation mechanisms, is the same in all classes of jets. Here we report on the discovery of a giant, highly-collimated jet from Sanduleak's star in the Large Magellanic Cloud (LMC). With a physical extent of 14 parsecs at the distance of the LMC, it represents the largest stellar jet ever discovered, and the first resolved stellar jet beyond the Milky Way. The kinematics and extreme chemical composition of the ejecta from Sanduleak's star bear strong resemblance with the low-velocity remnants of SN1987A and with the outer filaments of the most famous supernova progenitor candidate, i.e., $\eta$ Carin\ae{}. Moreover, the precise knowledge of the jet’s distance implies that it will be possible to derive accurate estimates of most of its physical properties. Sanduleak's bipolar outflow will thus become a crucial test-bed for future theoretical modeling of astrophysical jets. 
\end{abstract}

\keywords{ISM: jets and outflows --- binaries: symbiotic --- stars: individual (Sanduleak's star)}

\section{Introduction}
In 1977 N. Sanduleak reported on the discovery of a variable emission-line object in the direction of LMC (Sanduleak 1977). A comparison of objective-prism plates taken at the Cerro Tololo Inter-American Observatory in 1968, 1974 and 1975 showed strong emission in both the Balmer series and [OIII] $\lambda\lambda$5007, 4959\AA, and suspected variable emission at HeII $\lambda$4686\AA\ and [OIII] $\lambda$4363\AA. The suspected spectral variability, confirmed in later studies (Morgan et al. 1992), suggested that the source was ``some type of eruptive variable star rather than a planetary nebula'' (Sanduleak 1977).

The object (since then simply known as Sanduleak's star) was included by Allen (1980) among the symbiotic stars candidates in the LMC, in spite of some conflicting evidence that is still waiting for a suitable explanation, above all, the absence of any late-type stellar signatures in the optical spectra. IUE observations unveiled further peculiar spectral features, suggesting a possible similarity in the far-UV between Sanduleak's star, $\eta$ Carinae, and SN1987A (Michalitsianos et al. 1989). Earlier work found evidence of (i) episodic mass outflow, as indicated by the asymmetry of nitrogen line profiles; (ii) extreme departures from normal cosmic values of nitrogen relative to carbon and oxygen (i.e., N/C $\sim$ 150, N/O $\sim$ 70); and (iii) an electron temperature $T_e$ difficult to reconcile with pure photoionization models (Kafatos et al. 1983). \\ 
Nowadays, the symbiotic nature of Sanduleak's star appears, if not well-established, at least less contradictory: among the strongest proofs favouring a symbiotic classification there is the optical emission line spectrum, reminding of a dusty-type symbiotic star (Belczynski et al. 2000; Munari \& Zwitter 2002) and, even more constraining, the presence of the Raman bands at $\lambda\lambda$6825, 7082\AA, observed only from \textit{bona fide} symbiotic stars.

In this Letter, we report on the discovery of a giant, highly-collimated jet from Sanduleak's star in the LMC, obtained by using the Magellan Telescopes at the Las Campanas Observatory, Chile. In Sect. 2 we give some technical information about the observations, while in Sect. 3 we present the results of our analysis. Concluding remarks follow in Sect. 4.

\section{The Observations}
\begin{figure}
\epsscale{1}
\plotone{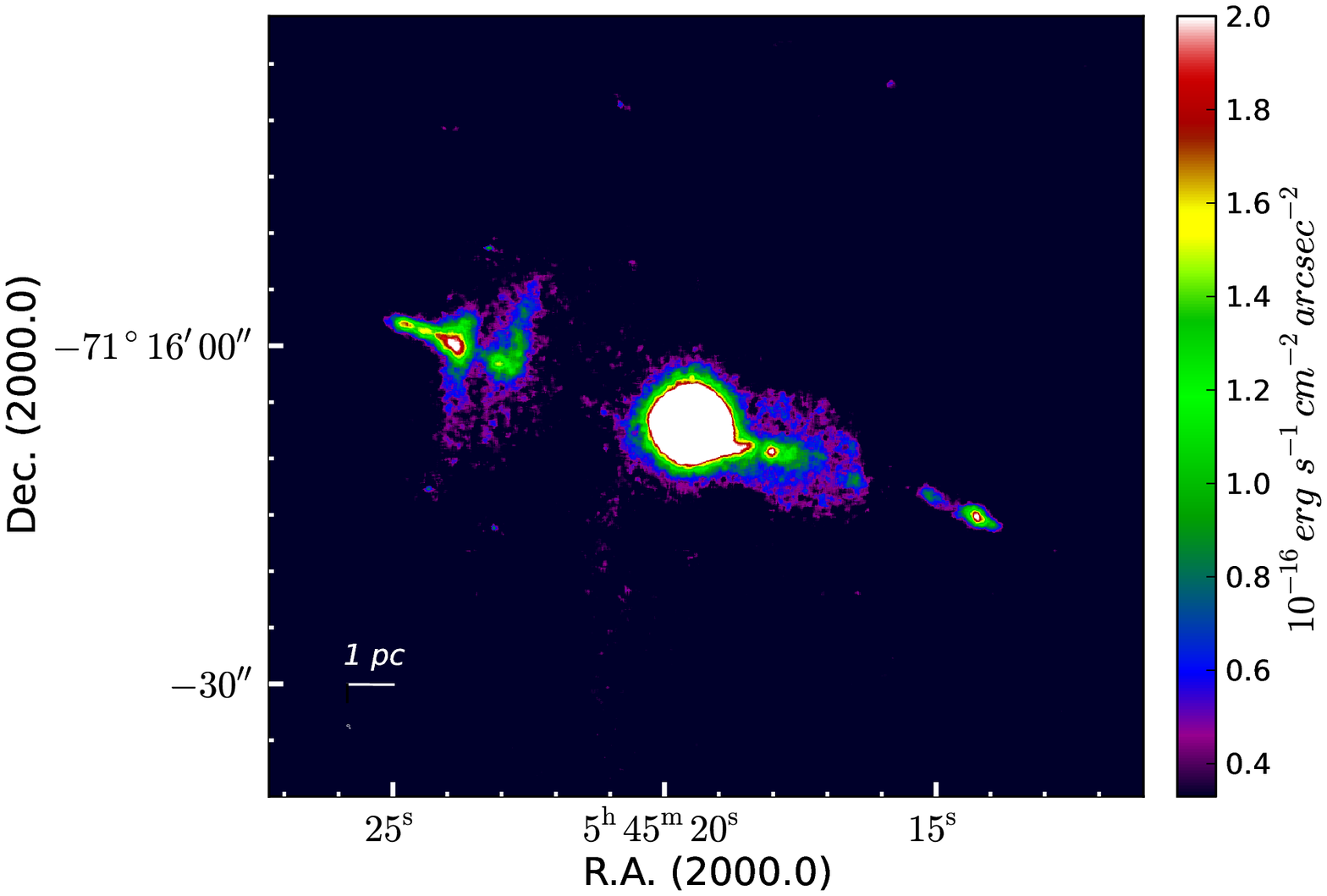}
\caption{$H\alpha$+[NII] emission line image of Sanduleak's star taken with the Raymond and Beverly Sackler Magellan Instant Camera (MagIC) at the Magellan-Clay Telescope, Las Campanas Observatory, Chile. The image is the result of the sum of three exposures of 900 sec each, through an interference filter which isolates the $H\alpha$6563+[NII]6548,6584 emission lines along with the underlying continuum. The equatorial coordinates and the absolute spatial scale of the image are also indicated. \label{fig1}}
\end{figure}

As part of a systematic observational campaign aimed at characterizing the symbiotic phenomenon outside the Milky Way, we have used the Las Campanas Observatory (LCO) Magellan Telescopes to obtain medium-resolution images and spectra of a representative sample of symbiotic stars in the Magellanic Clouds. For one of our survey object, Sanduleak's star, we have discovered a giant highly-collimated outflow, which we show in Fig. \ref{fig1}, where the equatorial coordinates and the absolute spatial scale of the image are also indicated.

The discovery image was taken with the \textit{Raymond and Beverly Sackler Magellan Instant Camera} (MagIC) at the Magellan-Clay Telescope on 2010 August 26, and it is the result of the sum of three exposures of 900 sec each, through an interference filter which isolates the $H\alpha$ 6563\AA+[NII]6548,6584\AA\ emission lines and the underlying continuum (effective wavelength $\lambda_{eff}$=6563\AA; full-width at half-maximum FWHM=50\AA). To obtain the pure emission line image we subtracted an appropriately scaled version of a 900 sec exposure taken through a narrow band filter which contains only continuum light ($\lambda_{eff}$=7130\AA, FWHM=50\AA).

In order to constrain the kinematics in the jet of Sanduleak's star, spectroscopy has been performed with the \textit{Inamori-Magellan Areal Camera and Spectrograph} (IMACS - Dressler et al. 2011) on Magellan-Baade Telescope. The spectrograph slit was 0.9 arcsecond wide, and it was aligned along the jet passing through the star at a Position Angle of PA=71\arcdeg. A 1200 sec exposure was recorded on 2010 October 23 with a 300 lines/mm grism blazed at 26.7\arcdeg, yielding a spectral sampling of 1.3\AA\ and a velocity resolution of 160 km/s in the spectral range 6300-9300\AA. The original spatial sampling is 0.2 arcsecond/pix, but the spectrum was rebinned along the spatial direction (0.6 arcsecond/pixel) to increase the signal to noise per resolution element. An additional 900 sec exposure was obtained in the same night with the 300 lines/mm grism blazed at 17.5\arcdeg, covering the wavelength range 4000-8000\AA\ with similar dispersion and resolution of the redder grism.

\section{Analysis and results}
\subsection{Imaging \label{imaging}}

The images obtained through the $H\alpha$+[NII] narrow-band filter clearly reveal a jet with an overall extent of 58 arcseconds. Using a distance modulus $\mu$=18.5 for the LMC (Schaefer 2008), this leads to a projected length of more than 14 pc. This makes it, to our knowledge, the \textit{largest bipolar stellar jet} ever observed, as well as the \textit{first resolved bipolar stellar jet beyond the Galaxy}. Adopting an upper limit of 2.4 arcseconds for the width of the jet, this corresponds to a lower limit of 24 for the aspect ratio of the highly-collimated structure. 

\begin{figure}
\epsscale{1}
\plotone{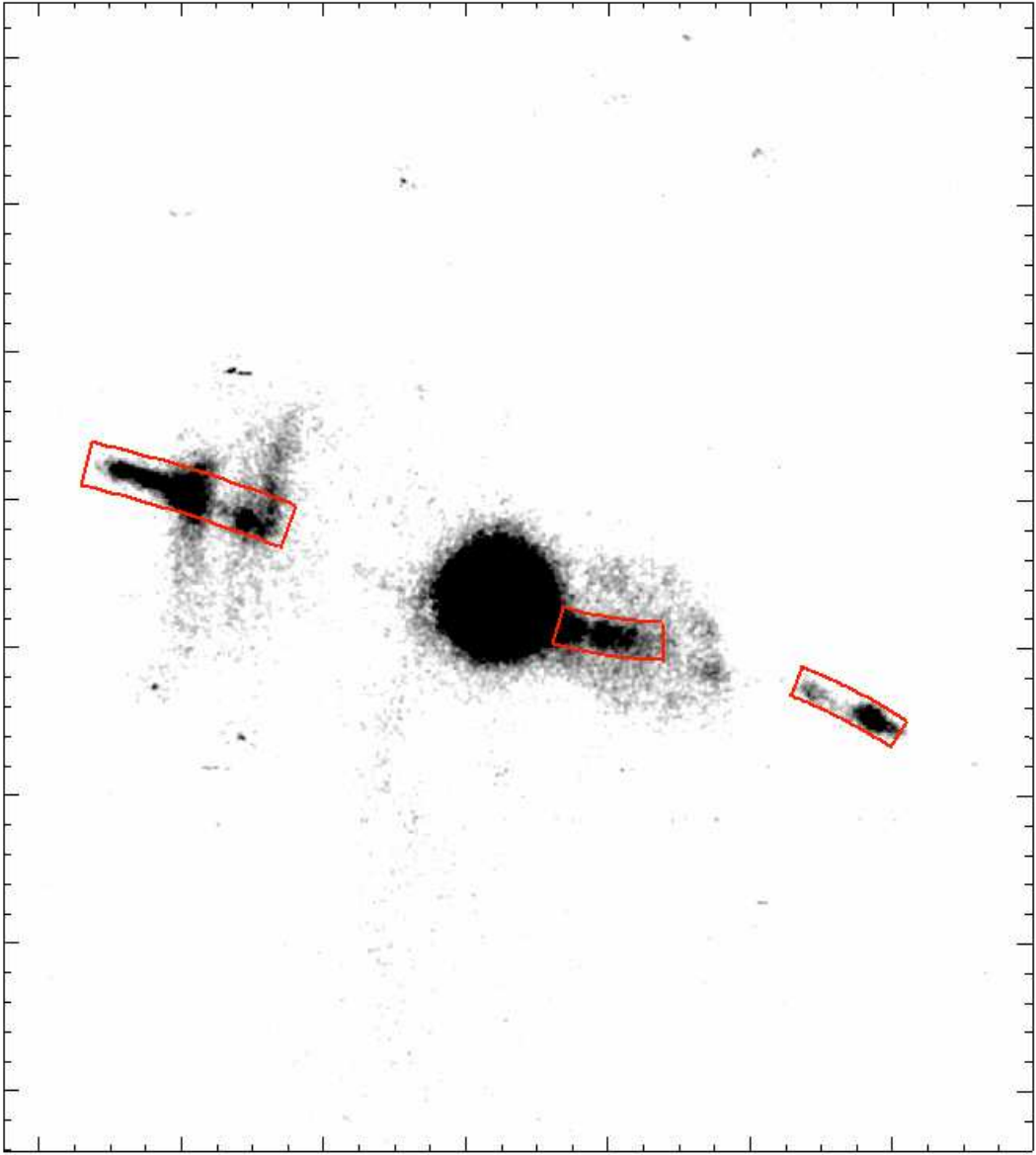}
\caption{Gray-scale version of Fig. \ref{fig1}. The red regions emphasize the jet curvature as traced by the consecutive knots, strong hints that the jet itself is likely precessing.    \label{fig1a}}
\end{figure}

The \textit{S-shaped} morphology of the entire complex hints that the jet is precessing (Fig. \ref{fig1a}). This behaviour is known to occur across a wide class of stellar sources (Karovska et al. 2010, Terquem et al. 1999). Furthermore, the opposite curvature displayed by the SW lobe at its extremes leads us to speculate that the precessional period is comparable to - or shorter than - the travel time taken for the gas to flow over the 7 pc of the SW lobe. 

\begin{figure}
\epsscale{1}
\plotone{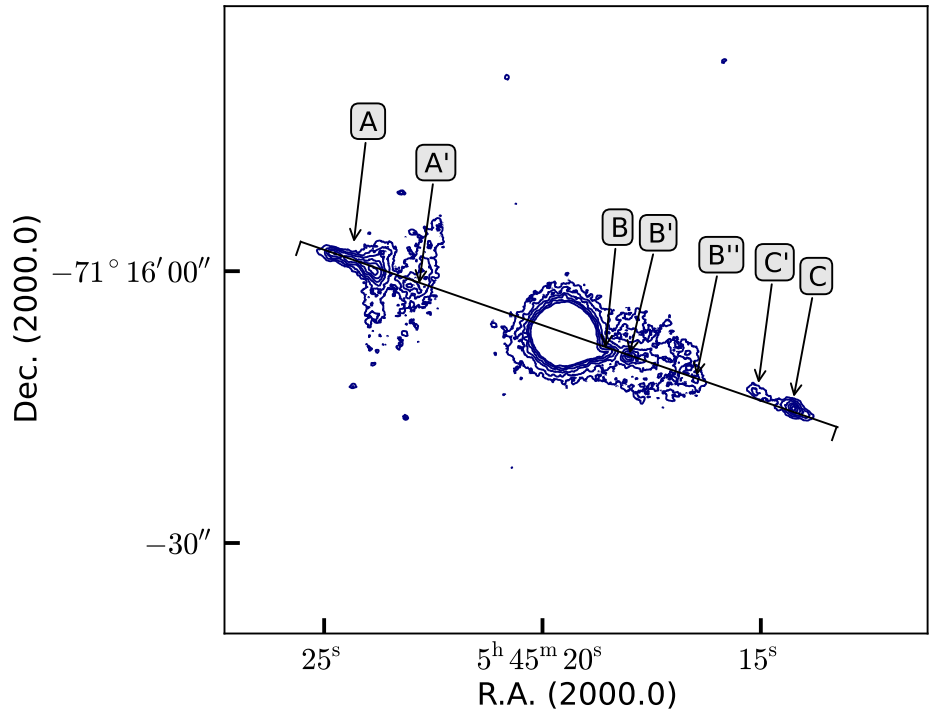}
\caption{Logarithmic contour map of the Hα+[NII] image of Sanduleak's star, with the same orientation and spatial scale as in Fig. \ref{fig1}. The structures discussed in the text are indicated with capital letters, while the solid line marks the slit position of the spectrum presented in Figs. \ref{fig3} and \ref{fig4}. \label{fig2}}
\end{figure}

Focusing on the small-scale structures, the jet complex is made up of a fairly regular and exceptionally symmetric pattern of knots and apparent bow-shocks (see Fig. \ref{fig2} for the nomenclature of the structures described hereafter). The most notable features are the opposite and highly symmetric pairs of “spear-like” knots marking the very edge of the flow (named as A and C in Fig. \ref{fig2}), which appear as its brightest regions in the $H\alpha$+[NII] image. Both located at 29 arcseconds from the central source, they display a distinguished \textit{V-shaped} configuration and have the remarkable size of 2 pc (knot A) and 1.5 pc (knot C), respectively.
Two broad arcs of emission perpendicular to the flow axis are clearly visible in the proximity of the NE knots (A and A'), extending for almost 5 pc. They bear  a close similarity with the feature N3-N6 visible in the HST images of the classic symbiotic star R Aqr (Paresce \& Hack 1994), even if in the latter case these structures measure only a few tens of AU. 
The SW lobe is characterized by a series of smaller, less collimated knots that seem to have cleared (C and C') or are still clearing (B, B' and B'') their way through an elongated nebula which, very close to the energy source and with a projected diameter of almost 3 pc, displays the characteristic of clumpy turbulence due to hydrodynamical (e.g., Rayleigh-Taylor) instabilities at the working surface. The morphological similarities with respect to analogous structures found in other classic symbiotic systems on vastly smaller scales (e.g., He2-104 -- Contini \& Formiggini 2001) is remarkable.

Observed integrated fluxes of the brightest knots are $1.94\times10^{-15}$, $6.24\times10^{-16}$, $2.16\times10^{-16}$, $5.81\times10^{-16}\ erg\ cm^{-2}\ s^{-1}$ for the knots A, A', C, and C', respectively. Their corresponding  $H\alpha$+[NII] luminosities are thus of a few 10$^{32}\ erg\ s^{-1}$, still several order of magnitude larger than knot luminosities in typical Herbig-Haro objects (Raga \& Noriega-Crespo 1998) and planetary nebulae (Gon\c{c}alves et al. 2009). The optical luminosity of the complex (assuming that $H\alpha$+[NII] dominate its emission and without taking into account the contribution of the central source) amounts to $3.5\times10^{33}\ erg\ s^{-1}$. 

\subsection{Spectroscopy}

\begin{figure}
\epsscale{1}
\plotone{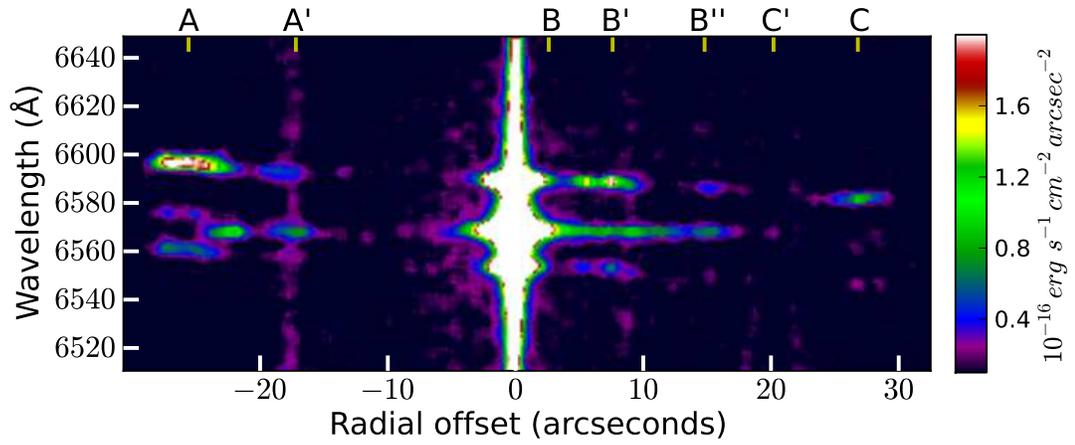}
\caption{Section of the spectrum of Sanduleak's star jet from the Inamori-Magellan Areal Camera and Spectrograph (IMACS - Dressler et al. 2011) on Magellan-Baade Telescope. The spectrograph slit was 0.9 arcsecond wide, and it was aligned along the jet passing through the star at a Position Angle of PA=71\arcdeg. \label{fig3}}
\end{figure}

\begin{figure}
\epsscale{1}
\plotone{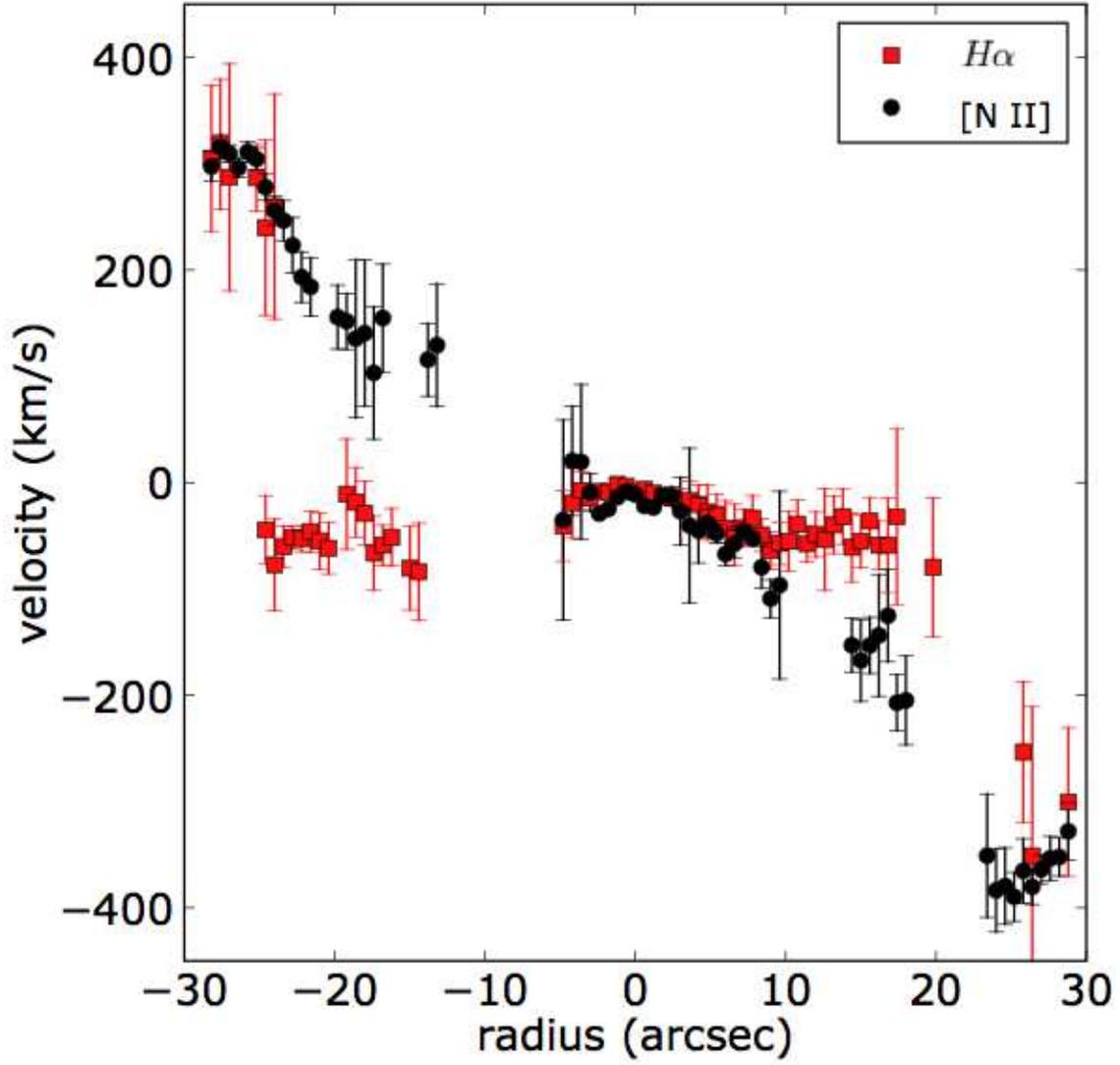}
\caption{Position-velocity plot of the Sanduleak's star jet. Radial velocities are quoted with respect to the systemic heliocentric velocity (270 km/s). \label{fig4}}
\end{figure}

The IMACS spectrum mentioned in Sect. 2 is finally presented in Fig. \ref{fig3}. \\
It shows [NII] tracing the global jet kinematics (with the NE lobe receding from, and the SW lobe moving toward us) which is characterized by a distinctive profile (Fig. \ref{fig4}) in which the flow radial velocity $v_{rad}$ increases linearly with the distance $r$ from the origin ($v_{rad} \propto r$). This behaviour is fairly common in proto-planetary nebulae (Corradi 2004, Huggins 2007) and eruptive stars (Weis et al. 1999). It is believed to indicate that the outflow is dictated by self-similar expansion, confirming the ballistic character of the knot trajectory and the explosive dynamics of its ejection (Dennis et al. 2008). The high degree of symmetry in the point-velocity distribution of the knots strongly suggests that the gas has been ejected in opposing directions at the same speed, giving weight to the hypothesis of recurrent outbursts from the central source at the origin of Sanduleak's star jet. Nova-like outbursts have already been invoked for explaining the appearance of supersonic knots and complex bipolar distributions in other sources (O'Connor et al. 2000, Livio \& Soker 2001).

In this ejection scenario, a few characteristic time scales can be derived. Because of its very large size ($\gtrsim14$ pc) and moderate radial velocity ($v_{rad}\lesssim400$km/s), the flow motion is likely to be almost orthogonal to our line of sight. If we assume a moderate inclination of the jet with respect to the plane of the sky ($\phi \lesssim 15\arcdeg$), the deprojected velocity would then amount to $v_{jet}\sim1500$ km/s, value that is consistent with the observations that typical velocity of jets in symbiotic stars is of the order of a few thousands km/s (Leedj\"{a}rv 2004). This result gives therefore weight to the hypothesis of a low-inclination system. With this in mind, we can now obtain an order-of-magnitude estimate of the jet kinematical age $t_{kin}$ by assuming as representative quantities \textit{i)} the largest observed radial velocity $V_{rad}=max\,|v_{rad}|$, and \textit{ii)} the jet size $s$: then, $t_{kin} = s\ V_{rad}^{-1}\ \tan \phi$. For small angles ($\phi \sim 15\arcdeg$) and for $V_{rad}\sim400$ km/s (as from Fig.\ref{fig4}), the derived kinematical age amounts to $\sim10^4$ years, which in turn provides an upper limit for the above mentioned precessional period (see Sect. \ref{imaging} and Fig. \ref{fig1a}). 
If one thus adopts the kinematical age of the farthest knots to be $\sim10^4$ years, a timescale for the ejection period of $\sim1000$ years is then found considering the average spacing between consecutive knots. This fact, combined with its faint optical magnitude ($V\sim17$), explains why there is no historical record of an earlier outburst from Sanduleak's star.

The jet seems also embedded in a diffuse ionized nebula: H$\alpha$ emission, particularly strong in the inner part of SW lobe but also visible in the NE one, shows a negative relative velocity of a few tens of km/s with respect to the systemic velocity. Within the inner 2 pc, both  H$\alpha$ and [NII] appear kinematically coupled, and display a distinctive wave-like modulation that is suggestive of a precessing emitting source. At larger distances, [NII] follows an increasing linear trend with velocity, while  H$\alpha$ reaches  negative velocities of 60-100 km/s up to 4.8 pc from the origin. This emission may be due to some kind of ionized stellar flow (even if the velocity is too high for considering a slow wind from a hidden giant companion, as in the dusty-type symbiotic star HM Sge - Muerset el al. 1991), that is shaping an expanding, perhaps axisymmetric nebula, like the one observed in M2-9 (Livio \& Soker 2001). 

A further H$\alpha$ component, this time clearly associated with [NII], emerges from the NE knots A and A': in this case, it seems reasonable to interpret it as stemming from the post-shock cooling region created by the interaction between the supersonic knots and the preexisting neutral hydrogen, collisionally excited before the atoms are ionized. The origin of this neutral material may date back to the time when the present outbursting star (probably a hot, luminous white dwarf of $T_{eff}\sim10^5$ K -- Muerset et al. 1996) was in the AGB superwind phase. Or it may be the result of ongoing episodes of intense mass-loss from an extincted giant companion, whose suggested mass-loss rate had been estimated at around $10^{-5}M_{\odot}$/year (Michalitsianos et al. 1989).\\

A careful examination of the long-slit spectrum across the jet (Fig. \ref{fig3}) reveals that the bulk of the knot emission comes from [NII], unusually strong (e.g., [NII]/H$\alpha\sim6$ in knot A, [NII]/H$\alpha\sim4$ in knot C), providing incontrovertible evidence of shock-ionized gas (Sabbadin \& D'Odorico 1976). No traces of [OIII] $\lambda\lambda$4959, 5007 and [OI] $\lambda$6300, nor of the [SII] $\lambda\lambda$6717-6731 doublet are detected from the knots, preventing us from giving any reliable estimate of the physical conditions (e.g., electron temperature and density) in the flow. Nonetheless, the upper limits in the line ratios that one can derive on the basis of these non-detections ([OIII]/H$\alpha \lesssim0.2$; [SII]/H$\alpha\lesssim0.2$) suggest a fairly extreme chemistry, in which nitrogen is strongly enhanced over oxygen, and where sulphur (a valuable tracer of low electron density in shocked environments) seems virtually absent. This peculiar composition, namely, the observed large nitrogen overabundance, is strikingly similar to what is seen in the ejecta of eruptive high-mass stars, such as the low-velocity remnants of SN1987A (Fransson et al. 1989) and the so-called \textit{S Condensation} of $\eta$ Carinae (Gull 2008). It may result from the ashes of the CNO cycle that have been exposed at the star's surface as a consequence of core-envelope mixing episodes during the AGB phase of the present WD. In this framework, we are then observing the nuclear processed material ejected during the recurrent outbursts that have characterized Sanduleak' star's recent history, a scenario that has been suggested also in the case of young symbiotic novae (Muerset \& Nussbaumer 1995). If this idea is correct, higher-quality data should be able to highlight spatial variation in the chemical abundance along the jet. In addition, the absence of any detectable sulfur emission (also barely visible in the spectra of $\eta$ Carinae outer filaments -- Smith \& Morse 2004) might be explained by the intrinsically lower sulfur content shown by evolved objects in the LMC with respect to the Milky Way (Bernard-Salas et al. 2008). Much longer exposures will be needed for recording the weak emission arising from these less abundant elements.

\section{Concluding remarks}
The true nature of Sanduleak's star remains a mystery, especially so in light of our discovery of the largest stellar jet to date. Despite the striking kinematical and chemical similarity with the ejecta of some high-mass eruptive stars, it seems unlikely that Sanduleak's star was as massive as, for example, $\eta$ Carinae. We note that the star’s infrared colors are similar to infrared colors of planetary nebulae and other dusty symbiotic objects in the LMC (Miszalski et al. 2011). Moreover, its monotonically fading long-term optical light curve is reminiscent of symbiotic novae that have undergone a recent thermonuclear outburst (Angeloni et al. in preparation, Muerset et al. 1994). Thus we support the idea that Sanduleak's star is a (yet outstanding) nova-like symbiotic, whose cool component is in all probability a post-AGB severely absorbed by an optically thick circumstellar envelope of dust and gas (Di Mille et al., in preparation). This suggestion sounds reasonable when considering that we already know a few galactic symbiotic stars in which the cool component is not visible in the optical because of the tremendous circumstellar extinction (e.g. H1-36 - Angeloni et al. 2007; HM Sge - Muerset et al. 1991, Angeloni et al. 2010); and that the advent of space-based infrared observatories (such as AKARI and SPITZER) has unveiled a number of AGB and post-AGB stars completely invisible in the optical range but extremely bright in the infrared {(Bunzel et al. 2009; Garc\'{\i}a-Hern\'{a}ndez et al. 2009).}\\

The size and luminosity of the individual knots in the jet; their high collimation, maintained at distances beyond 7 pc from the central source despite a not-relativistic flow speed; the overall size of the jet itself: all this represents observational evidence that challenges our current understanding of astrophysical jets. And that so far has not been dealt with in theoretical models and numerical simulations simply because nobody would have ever bet that such a giant stellar jet could exist. In this respect, and whatever its origin, we think that the Sanduleak bipolar outflow is destinated to turn into a crucial test-bed for future modeling of astrophysical jets. 

\acknowledgments
Support for RA is provided by Proyecto Fondecyt \#3100029 and by the Ministry for the Economy, Development, and Tourism's Programa Iniciativa Cient\'{i}fica Milenio through grant P07-021-F, awarded to The Milky Way Millennium Nucleus. FDM acknowledges the support of a Magellan Fellowship from Astronomy Australia Limited, and administered by the Australian Astronomical Observatory. JBH is supported by a Federation Fellowship from the Australian Research Council. This paper includes data gathered with the 6.5 meter Magellan Telescopes located at the Las Campanas Observatory, Chile. Australian access to the Magellan Telescopes was supported through the National Collaborative Research Infrastructure Strategy of the Australian Federal Government. The authors would like to thank the anonymous referee, and Mark Phillips, for commenting on the manuscript, and acknowledge the Las Campanas staff.

{\it Facilities:} \facility{LCO (Magellan Telescopes)}.

\end{document}